\documentclass[11pt]{article}
\usepackage{moriond,epsfig,verbatim,subfigure,hhline}
\usepackage[english]{babel}

\begin{document}

\newcommand{\dub}[1]{\textbf{[??? #1 ???]}}

\vspace*{4cm}

\title{DISPERSION IN OPTICAL FIBERS AND TIMING FOR PARTICLE IDENTIFICATION}

\author{FABIO PALEARI}

\address{Dipartimento di Scienze Chimiche, Fisiche, Matematiche, Universit\`a dell'Insubria, via Valleggio 11,\\
20111 Como, Italy; INFN sezione di Milano, piazza della Scienza 3,
20126 Milano.}
\maketitle
\abstracts{In the framework of the TOF Wall laser calibration system
of the HARP experiment, a study of time dispersion
properties of mono-mode and multi-mode optical fibers in the
green band (532\,nm) has been carried out. Dispersion less than 4\,ps/m
has been obtained with $\approx$10\,$\mu$m core diameter fibers.}

In this report a study  on the time dispersion properties of different
fiber types  is presented. In  particular, IR mono-mode fibers  in the
green  band has  been analyzed:  in  this range  of wavelengths  these
fibers work as  ``not-so-many-mode'' fibers.  The aim of  the study is
the design of the laser calibration system\,\cite{ArticoloHarp} of the
TOF~(Time Of Flight)~Wall\,\footnote{The  TOF Wall  is  a (7.2$\times$3)\,m${}^2$  wall
made   of  39~(250   or   180$\times$20$\times$2.5)\,cm${}^3$  plastic
scintillator counters  (Bicron BC408)  read at each  end by  a Philips
XP2020  photomultiplier.  With  an  overall resolution  of 160\,ps,  a
3$\sigma$  separation between  protons and  pions can be achieved  up to
momenta of 3\,GeV/c.}
of the  PS214-HARP experiment\,\cite{Hprop}, an  hadronic spectrometer
running  on the  T9  beam at  the CERN~24\,GeV~Proton~Synchrotron,  in
which TOF measurements are used for particle identification.

High precision TOF  measurements, required by particle
identification,  need a fast  and accurate  calibration system  (down to
100\,ps or less), to keep under control possible drifts in  the time response of
the detector arrising from the  different experimental conditions.  A suitable
solution, for  scintillating counters, is  to use  a short
pulsed  laser  source, in  the  visible  range,  to simulate  particle
interactions in the counters. The light is distributed to the counters
by a  bundle of optical  fibers then, precise timing requires  that minimal
modification  of  the pulse  leading  edge take place during the  propagation
through the fibers.

Mono-mode fibers \cite{FuPho}
%
%
provide negligible dispersion at a given wavelength for distances of some
kilometers, but they need  very accurate injection devices, because of
the very limited  acceptance angle and core dimensions  (few times the
optimized light wavelength); moreover, in-fiber division systems are
necessary  to obtain  a  multi-fiber bundle.   On  the other  side,
multi-mode fibers  do not present  injection difficulties but,  due to
modal dispersion up  to 30\,ps/m, they are useless  for precise timing
over distances larger than few tenth of centimeters.

\begin{figure}[t!] 
\scalebox{.9}{
\begin{minipage}{.6\textwidth}
\begin{tabular}{c}
\psfig{figure=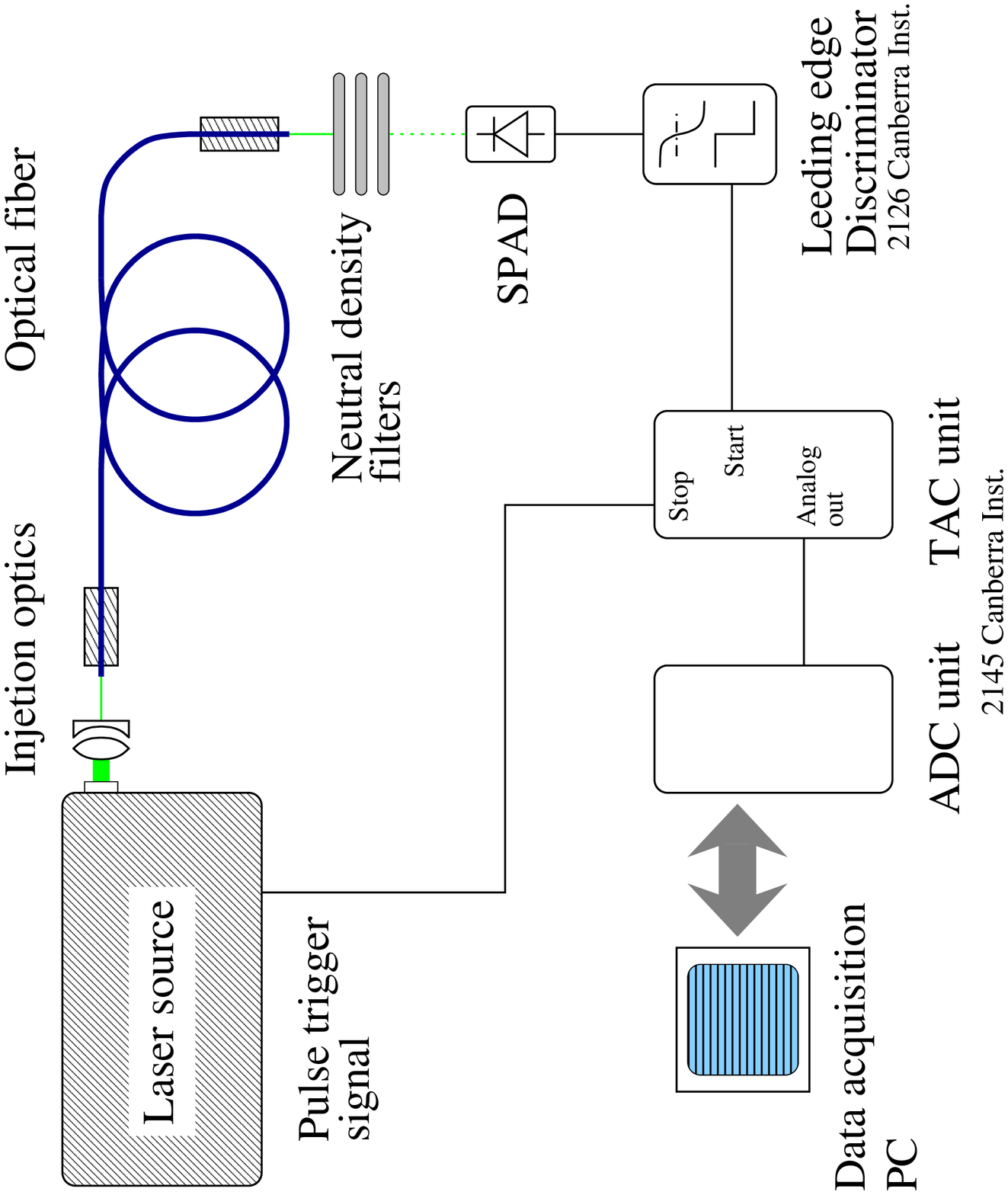,height=10cm,angle=-90}\\      {\small
Figure 1: Setup for the dispersion measure.}
\end{tabular}
\end{minipage}
\begin{minipage}{.4\textwidth}
\begin{tabular}{l}
{\small Table 1: Results for dispersion (D)}\\
{\small ~~~~~~~~~~~~of the fibers.}\\ \\
\begin{tabular}{|c|c|}\hline
\multicolumn{2}{|l|}{Fiber}\\
Type&\textbf{D\,(ps/m)}\\ \hline 
\multicolumn{2}{|l|}{CERAM OPTEC 100/110}\\
multi-mode&\textbf{14.9}\\\hline
\multicolumn{2}{|l|}{Shuner  FiberOptics  90/125}\\
multi-mode&\textbf{8.74}\\ \hline
\multicolumn{2}{|l|}{FOS SMR-R}\\
IR mono-mode&\textbf{3.5}\\ \hline
\multicolumn{2}{|l|}{Corning SMF-18}\\
IR mono-mode&\textbf{3.6}\\ \hline
\end{tabular}
\end{tabular}
\end{minipage}
\label{setup}
}
\end{figure}

We  used the second  harmonic output  of a  continous-wave
mode-locked\,\cite{FuPho} Nd:VAN  laser (Time  Bandwidth GE 100-1064
VAN) to   provide a 9\,ps pulse
at 532\,nm  (green) and we measured the  pulse shape with  a single photon
timing technique\,\footnote{In the low intensity
limit  the  time  distribution  of  the  clicks  of  the  detector  is
proportional to  the real probability  distribution of the  photons in
the pulse.}. The  light  detector is a  single-photon  avalanche  photodiode
(SPAD),  which is  read   with  a  TAC-ADC  system  with  1.98\,$\mathrm{ps/Ch}$
resolution.  The setup is outlined in figure 1.

To determine  the pulse  time response of  the whole system, at first we
measured  the time distribution of the photon in the laser pulse,
without optical  fiber  (obtaining
40\,ps~FWHM), then  we sent the  laser pulse through the fiber
repeating the measure. We  subtracted in  quadrature  the RMS  of the  two
measurements and divided by the sample lengths. The results are given in table 1.

In conclusion, we find that IR mono-mode fibers used with green light
can fulfill timing  requirements for particle identification detectors
up to  several meter distances. In the HARP TOF Wall, we use
Corning~SMF-28~fibers, obtaining  an overall sensitivity of 70\,ps in  the time
response drifts.
\section*{Acknowledgments}
Dispersion measures was made in fruitfull collaboration with the group of Prof.
Sottocornola Spinelli (Universit\`a dell'Insubria).

\end{document}